\documentclass[twocolumn,english,showpacs,preprintnumbers,amsmath,amssymb]{revtex4}
\usepackage[T1]{fontenc}
\usepackage[latin1]{inputenc}
\usepackage{graphicx}

\makeatletter

\providecommand{\tabularnewline}{\\}

\usepackage{graphicx}
\usepackage{dcolumn}
\usepackage{bm}

\usepackage{babel}
\makeatother
\begin{document}

\title{Crossover behavior in failure avalanches}

\author{Srutarshi Pradhan}

\email{pradhan.srutarshi@ntnu.no}

\author{Alex Hansen}

\email{alex.hansen@ntnu.no}

\author{Per C. Hemmer}

\email{per.hemmer@ntnu.no}

\affiliation{Department of Physics, Norwegian University of Science and Technology,
N--7491 Trondheim, Norway}

\begin{abstract}
Composite materials, with statistically distributed threshold for
breakdown of individual elements, are considered. During the failure
process of such materials under external stress (load or voltage),
avalanches consisting of simultaneous rupture of several elements
occur, with a distribution $D(\Delta)$ of the magnitude $\Delta$
of such avalanches. The distribution is typically a power law $D(\Delta)\propto\Delta^{-\xi}$.
For the systems we study here, a crossover behavior is seen between
two power laws, with a small exponent $\xi$ in the vicinity of complete
breakdown and a larger exponent $\xi$ for failures away from the
breakdown point. We demonstrate this analytically for bundles of many
fibers where the load is uniformly distributed among the surviving
fibers. In this case $\xi=3/2$ near the breakdown point and $\xi=5/2$
away from it. The latter is known to be the generic behavior. This
crossover is a signal of imminent catastrophic failure of the material.
Near the breakdown point, avalanche statistics show nontrivial finite
size scaling. We observe similar crossover behavior in a network of
electric fuses, and find $\xi=2$ near the catastrophic failure and
$\xi=3$ away from it. For this fuse model power dissipation avalanches
show a similar crossover near breakdown. 
\end{abstract}
\maketitle

\section{Introduction}

Burst avalanches play an important role in characterising the fracture-failure
phenomena \cite{book-1,book-2,book-3,book-4}. When a weak element
in a loaded material fails, the increased stress on the remaining
elements may cause further failures, and thereby give a failure avalanche
in which several elements fail simultaneously. With further increase
in the load new avalanches occur. The statistics of such avalanches
during the entire failure process explore the nature of correlations
developed within the system. From the experimental point of view,
failure avalanches are the only measurable quantity during the fracture-failure
process of composite materials \cite{AE-1,AE-2,AE-3}. Under quasi-static
loading the system, with some internal load redistribution mechanism,
gradually approaches the global failure point. Such damage and fracture
of materials are of immense interest due to their economic and human
costs. Therefore, a fundamental challenge is to find methods for providing
signals that warn of imminent global failure. This is of uttermost
importance in, e.g., the diamond mining industry where sudden failure
of a mine is always catastrophic. These mines are under continuous
acoustic surveillance, but at present there is no meaningful acoustic
signature of imminent disaster. The same type of question is of course
central to earthquake prediction \cite{book-2,book-3,book-4}.

In this paper we will study crossover behavior of failure avalanches
in the context of two very different models where the system gradually
approaches global failure through several intermediate failure events.
We find that if a histogram of the number of elements failing simultaneously
is recorded, it follows a power law with an exponent that crosses
over from one value to a very different value when the system is close
to global failure. This crossover is, then, the signature of imminent
breakdown. The first system studied here is a bundle of many fibers
\cite{FT,Dan} with stochastically distributed fiber strengths. This
model is sufficiently simple that an analytic treatment is feasible
\cite{HH-92,HH-94,JV-97,SPB-02,PSB-03,PH-04,PHH-05,KHH-97}. The second
system is a fuse model \cite{book-1}, a two-dimensional lattice in
which the bonds are fuses, i.e., ohmic resistors with stochastically
distributed threshold values. This model must be analyzed numerically.
Both models exhibit similar crossovers as signal of imminent breakdown.

The paper is organized as follows. In Section II  we present numerical
evidence for the crossover in the fiber bundle model, backed up by
analytic derivations. We pay particular attention to the burst properties
just before complete breakdown. Cascading failures in a fuse model
is the theme of Section III.

\section{The fiber bundle model}

\subsection{Numerical evidence}

A bundle of many fibers with stochastically distributed fiber strengths,
and clamped at both ends, is a much-studied model \cite{FT,Dan,HH-92,HH-94,KHH-97,JV-97,SPB-02,PSB-03,PH-04,PHH-05}
for failure avalanches. In its classical version, a ruptured fiber
carries no load and the increased stresses caused by a failed element
are shared equally by all the surviving fibers. The maximal loads
$x_{n}$ that the fibers $n=1,2,\ldots,N$ are able to carry are picked
independently with a probability density $p(x)$: \begin{equation}
\mbox{Prob}(x\leq x_{n}<x+dx)=p(x)\; dx.\end{equation}
 A main result for this model is that under mild restrictions on the
fiber strength distribution the expected number $D(\Delta)$ of burst
avalanches in which $\Delta$ fibers fail simultaneously is governed
by a universal power law \cite{HH-92}\begin{equation}
D(\Delta)\propto\Delta^{-\xi}\label{powerlaw}\end{equation}
 for large $\Delta$, with $\xi=5/2$. However, we will show that
when the whole bundle is close to breaking down the exponent crosses
over to a lower value. Such a complete breakdown can be estimated
as follows: The force $F(x)$ that the bundle is able to withstand
when all fibers with strengths less than $x$ have ruptured, is proportional
to the number of surviving fibers times the strength, \begin{equation}
F(x)=Nx\; Q(x),\label{F}\end{equation}
 where \begin{equation}
Q(x)=\int_{x}^{\infty}p(x)\, dx\end{equation}
 is the expected fraction of fibers with strengths exceeding $x$.
As an example, assume the threshold distribution $p(x)$ to be uniform
in an interval $(x_{0},x_{m})$, \begin{equation}
p(x)=\left\{ \begin{array}{cl}
(x_{m}-x_{0})^{-1} & \mbox{for }x_{0}\leq x\leq x_{m}\\
0 & \mbox{otherwise}.\end{array}\right.\label{uniform}\end{equation}
 In this case we obtain \begin{equation}
F(x)=N{\displaystyle \frac{x(x_{m}-x)}{x_{m}-x_{0}}},\end{equation}
 which has a maximum at $x_{c}=x_{m}/2$. In general we call $F(x)$
the \textit{average} force and the value corresponding to the maximum
of $F(x)$ the \textit{critical}\ threshold value $x_{c}$. If $F(x)$
given by (\ref{F}) were the actual force, the bundle would break
down when $x$ reaches the value $x_{c}$. By the existence of fluctuations,
however, the maximum value of the force may actually occur at a slightly
different (probably higher) value of $x$.

\begin{center}\includegraphics[%
  width=2.4in,
  height=2.1in]{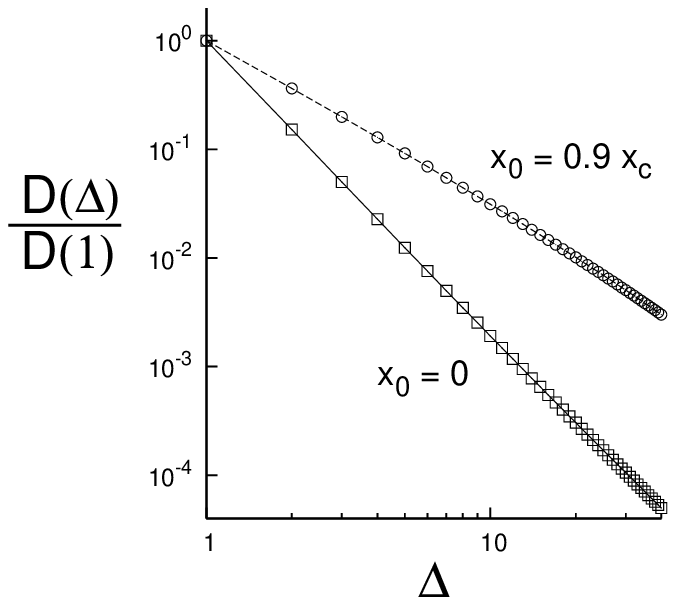}\end{center}

{\footnotesize FIG. 1.} {\footnotesize The distribution of bursts
for the strength distribution (\ref{uniform}) with $x_{0}=0$ and
$x_{0}=0.9x_{c}$. The figure is based on $50000$ samples with $N=10^{6}$
fibers.}{\footnotesize \par}

\vskip.1in

We want to study burst avalanches when the weakest fiber $x_{0}$,
is close to the critical value $x_{c}$. In Fig.\ 1 we show results
for $D(\Delta)$ for the uniform distribution with $x_{0}=0.9\, x_{c}$.
For comparison, simulation results with $x_{0}=0$ are shown. In both
cases $D(\Delta)$ shows a power law decay, apparently with an exponent
$\xi=3/2$ for $x_{0}=0.9\, x_{c}$ in contrast to the standard exponent
$\xi=5/2$ for the $x_{0}=0$ case.

In Fig.\ 2 we show that the same two exponents appear for a much
more concentrated threshold distribution, the Weibull distribution.
We will in the following explain the results as a crossover phenomenon.

\begin{center}\includegraphics[%
  width=2.4in,
  height=2.1in]{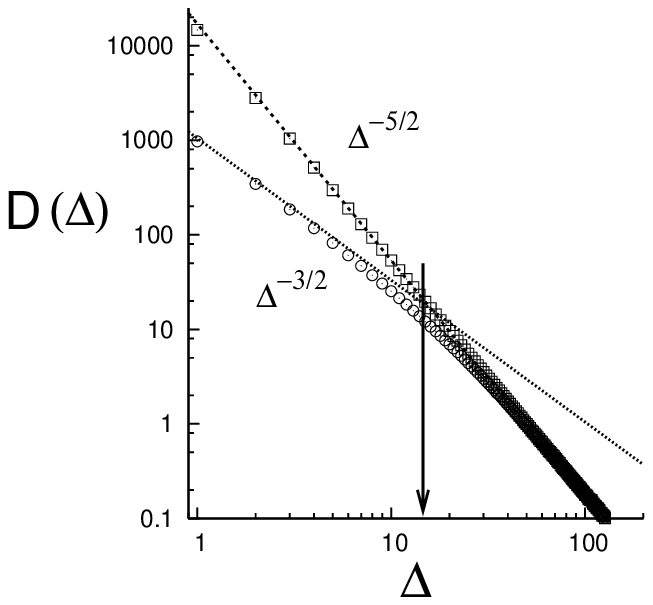}\end{center}

{\footnotesize FIG. 2.} {\footnotesize The distribution of bursts
for the Weibull distribution $Q(x)=\exp(-(x-1)^{10})$, where $1\leq x\leq\infty$.
Again results for the two cases $x_{0}=1$ (squares) and $x_{0}=1.7$
(circles) are displayed ($x_{c}=1.72858$ for this distribution).
The figure is based on $50000$ samples with $N=10^{6}$ fibers and
the arrow locates the crossover point $\Delta_{c}\simeq14.6$.}\\

\subsection{Analytical treatment of the crossover}

For a bundle of many fibers the number of bursts of length $\Delta$
is given by \cite{HH-92}

\begin{flushleft}{\footnotesize \begin{equation}
\frac{D(\Delta)}{N}={\displaystyle \frac{\Delta^{\Delta-1}e^{-\Delta}}{\Delta!}\int_{0}^{x_{c}}p(x)r(x)[1-r(x)]^{\Delta-1}\exp\left[\Delta\, r(x)\right]dx,}\label{D}\end{equation}
} where \begin{equation}
r(x)=1-\frac{x\, p(x)}{Q(x)}=\frac{1}{Q(x)}\;\frac{d}{dx}\left[x\, Q(x)\right].\label{eq:8}\end{equation}
\end{flushleft}

From the last expression we see that $r(x)$ vanishes at the point
$x_{c}$ where the average force expression (\ref{F}) is maximal.
If we have a situation in which the weakest fiber has its threshold
$x_{0}$ just a little below the critical value $x_{c}$ the contribution
to the integral in the expression (\ref{D}) for the burst distribution
will come from a small neighborhood of $x_{c}$. Since $r(x)$ vanishes
at $x_{c}$ it is small here, and we may in this narrow interval approximate
the $\Delta$-dependent factors in (\ref{D}) as follows

{\footnotesize \begin{eqnarray}
(1-r)^{\Delta}\, e^{\Delta\, r} & = & \exp\left[\Delta(\ln(1-r)+r)\right]\nonumber \\
 & = & \exp[-\Delta(r^{2}/2+{\mathcal{O}}(r^{3}))]\approx\exp\left[-\Delta r(x)^{2}/2\right]\label{D3}\end{eqnarray}
}{\footnotesize \par}

\noindent We also have \begin{equation}
r(x)\approx r'(x_{c})(x-x_{c}).\end{equation}
 Inserting everything into Eq.\ (\ref{D}), we obtain to dominating
order {\scriptsize \begin{eqnarray}
\frac{D(\Delta)}{N} & = & \frac{\Delta^{\Delta-1}\, e^{-\Delta}}{\Delta!}\int_{x_{0}}^{x_{c}}p(x_{c})\; r'(x_{c})(x-x_{c})e^{-\Delta\, r'(x_{c})^{2}(x-x_{c})^{2}/2}\; dx\nonumber \\
 & = & \frac{\Delta^{\Delta-2}\, e^{-\Delta}p(x_{c})}{\left|r'(x_{c})\right|\Delta!}\left[e^{-\Delta\, r'(x_{c})^{2}(x-x_{c})^{2}/2}\right]_{x_{0}}^{x_{c}}\nonumber \\
 & = & \frac{\Delta^{\Delta-2}\, e^{-\Delta}}{\Delta!}\frac{p(x_{c})}{\left|r'(x_{c})\right|}\left[1-e^{-\Delta/\Delta_{c}}\right],\label{D3}\end{eqnarray}
} with \begin{equation}
\Delta_{c}=\frac{2}{r'(x_{c})^{2}(x_{c}-x_{0})^{2}}.\label{Dc}\end{equation}

By use of the Stirling approximation $\Delta!\simeq\Delta^{\Delta}e^{-\Delta}\sqrt{2\pi\Delta}$
-- a reasonable approximation even for small $\Delta$, the burst
distribution (\ref{D3}) may be written as \begin{equation}
\frac{D(\Delta)}{N}=C\Delta^{-5/2}\left(1-e^{-\Delta/\Delta_{c}}\right),\label{D2}\end{equation}
 with a nonzero constant \begin{equation}
C=(2\pi)^{-1/2}p(x_{c})/\left|r'(x_{c})\right|.\end{equation}
 We see from (\ref{D2}) that there is a crossover at a burst length
around $\Delta_{c}$, so that \begin{equation}
\frac{D(\Delta)}{N}\propto\left\{ \begin{array}{cl}
\Delta^{-3/2} & \mbox{ for }\Delta\ll\Delta_{c}\\
\Delta^{-5/2} & \mbox{ for }\Delta\gg\Delta_{c}\end{array}\right.\end{equation}

\begin{center}\includegraphics[%
  width=2.4in,
  height=2.1in]{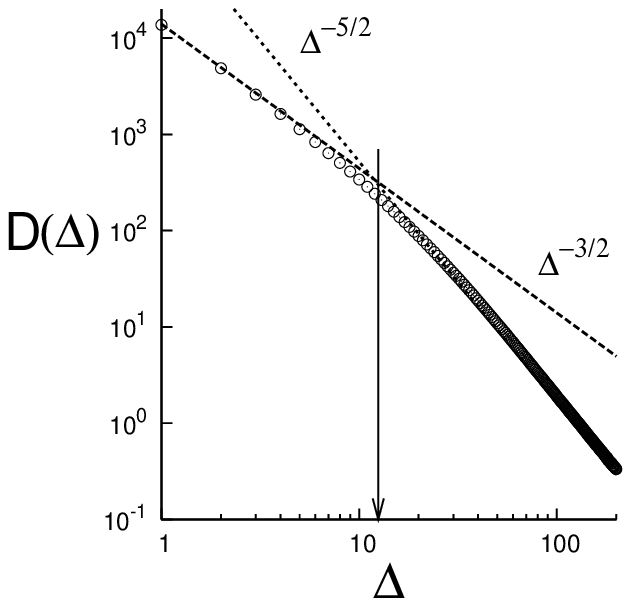}\end{center}

{\footnotesize FIG. 3.} {\footnotesize The distribution of bursts
for the uniform threshold distribution (\ref{uniform}) with $x_{0}=0.80x_{\textrm{c}}$.The figure is based on $50000$ samples with $N=10^{6}$ fibers. The
straight lines represent two different power laws, and the arrow locates
the crossover point $\Delta_{c}\simeq12.5$.}{\footnotesize \par}
\vskip.2in
 We have thus shown the existence of a crossover from the generic
asymptotic behavior $D\propto\Delta^{-5/2}$ to the power law $D\propto\Delta^{-3/2}$ near criticality, i.e., near global breakdown. 
The fact that there may be a different  burst distribution exponent
 near breakdown has been noted by Sornette (see Ref.1 and references therein), 
and observed  by Zapperi et al [19] for a fuse model.
The crossover is a
universal phenomenon, independent of the threshold distribution $p(x)$.
In addition we have located where the crossover takes place.

\vskip.1in

For the uniform distribution $\Delta_{c}=(1-x_{0}/x_{c})^{-2}/2$,
so for $x_{0}=0.9\, x_{c}$, we have $\Delta_{c}=50$. The final asymptotic
behavior is therefore not visible in Fig.\ 1. The crossover is seen
better for $x_{0}=0.8x_{\textrm{c}}$, as in Fig.\ 3. Now a crossover
is clearly observed near $\Delta=\Delta_{c}=12.5$, as expected.

The simulation results shown in the figures are based on \textit{averaging}
over a large number of fiber bundles with moderate $N$. For applications
it is important that crossover signals are seen also in a single sample.
We show in Fig. 4 that equally clear power laws are seen for a \textit{single}
fiber bundle when $N$ is large.

\begin{center}\includegraphics[%
  width=2.4in,
  height=2.1in]{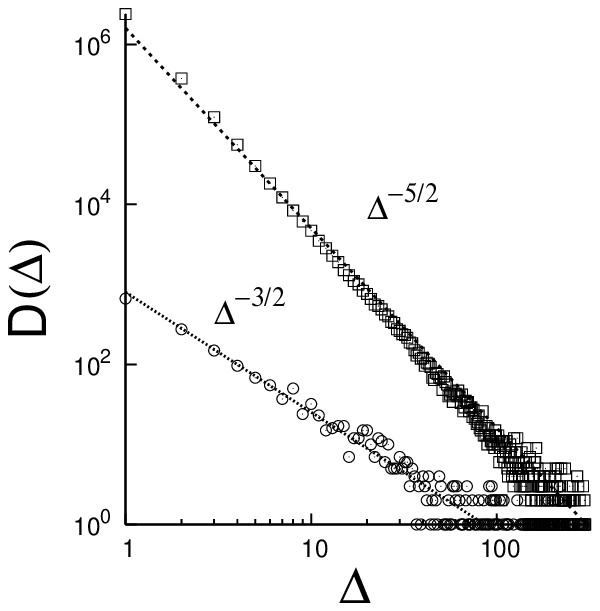}\end{center}

{\footnotesize FIG. 4.} {\footnotesize Avalanche distribution for
the uniform threshold distribution (\ref{uniform}) for a single fiber
bundle with $10^{7}$ fiber: all avalanches (squares) and avalanches
near the critical point (circles). Dotted straight lines are the best
fits to the power laws.}\\

\subsection{Sampling a finite interval}
Above we have explained the crossover in burst distributions in which all bursts up to complete breakdown are counted. For the purpose of finding signals of imminent global failure one must, of course, determine burst distributions short of complete breakdown. Consequently we are interested in sampling a finite interval $(x_0,x_f)$,
with $x_f<x_c$. When the interval is in the neighborhood of $x_c$ we have, as in Eq.(11)
\begin{eqnarray}
\frac{D(\Delta)}{N} &\simeq & \frac{\Delta^{\Delta-2}e^{-\Delta}p(x_c)}{|r'(x_c)|\Delta !} \left[e^{-r'(x_c)^2(x-x_c)^2\Delta/2}\right]_{x_0}^{x_f}\\
  &\simeq& C\Delta^{-5/2} \left(e^{- \Delta (x_c-x_f)^2/a}-e^{-\Delta (x_c-x_0)^2/a}\right),\;\;\;\;\;\;\;\;\;
\end{eqnarray}
with $a=2/r'(x_c)^2$.

This shows a crossover:
\begin{equation}
\frac{D(\Delta)}{N} = \left\{ \begin{array}{ll}
\widetilde{C} \; \Delta^{-3/2}& \mbox{ for } \Delta \ll a/(x_c-x_0)^2 \\
C \Delta^{-5/2} & \mbox{ for } a/(x_c-x_0)^2 \ll \Delta \ll a/(x_c-x_f)^2,
\end{array} \right.
\end{equation}
with a final exponential behavior when $\Delta \gg a/(x_c-x_f)^2$.
Here $\widetilde{C}= C a^{-1} \left[(x_c-x_f)^2-(x_c-x_0)^2\right]$.

The  $3/2$ power law  will be seen only when the beginning of the interval, $x_0$, is close
enough to the critical value $x_c$ to create a sizeable range of bursts obeying this power law. Observing the $3/2$ power law is therefore a signal of imminent system breakdown.

\subsection{Burst avalanches at criticality}

Precisely at criticality $(x_{0}=x_{c})$ we have $\Delta_{c}=\infty$,
and consequently the $\xi=5/2$ power law is no longer present. We
will now argue, using a random walk representation, that at criticality
the burst distribution follows a 3/2 power law. The load on the bundle
when the $k$th fiber with strength $x_{k}$ is about to fail is proportional
to \begin{equation}
F_{k}=x_{k}(N-k+1).\end{equation}
 The expectation value of this is the average force equation (\ref{F}).
At criticality the $F$ is, on the average, stationary. It is, however,
the \textit{fluctuations} of this load that now determines the size
of the bursts. It has been shown \cite{HH-94} that the probability
$\rho(f)\, df$ that the difference $F_{k+1}-F_{k}$ lies in the interval
$(f,f+df)$ is given by \begin{equation}
\rho(f)=\left\{ \begin{array}{ll}
\frac{1-r(x_{k})}{x_{k}}e^{-(1-r(x_{k}))(1+f/x_{k})} & \mbox{ for }f\geq-x_{k}\\
0 & \mbox{ for }f<-x_{k}\end{array}\right.,\end{equation}
 where $r(x)$ is given by Eq. \ref{eq:8}. At criticality $r=0$,
resulting in \begin{equation}
\rho_{c}(f)=\left\{ \begin{array}{ll}
x_{c}^{-1}\, e^{-1}\, e^{-f/x_{c}} & \mbox{ for }f\geq-x_{c}\\
0 & \mbox{ for }f<-x_{c}\end{array}\right.\label{eq: 18}\end{equation}
 This can be considered as the step probability in a random walk.
The random walk is unsymmetrical, but \textit{unbiased}, $\langle f\rangle=0$,
as it should be at criticality.

A first burst of size $\Delta$ corresponds to a random walk in which
the position after each of the first $\Delta-1$ steps is lower than
the starting point, but after step no. $\Delta$ the position of the
walker exceeds the starting point. The probability of this equals 

\vskip.1in

\begin{flushleft}{\small $\mbox{Prob}(\Delta)=\int_{-x}^{0}\rho(f_{1})df_{1}\int_{-x}^{-f_{1}}\rho(f_{2})df_{2}\int_{-x}^{-f_{1}-f_{2}}\rho(f_{3})df_{3}\ldots$}\end{flushleft}{\small \par}

\vskip -.2in

{\footnotesize \begin{equation}
..\int_{-x}^{-f_{1}-f_{2}...-f_{\Delta-2}}\rho(f_{\Delta-1})df_{\Delta-1}\int_{-f_{1}-f_{2}...-f_{{\Delta-1}}}^{\infty}\rho(f_{\Delta})\, df_{\Delta}.\label{int}\end{equation}
}{\footnotesize \par}

\vskip.2in

\noindent The last integral is easy. By means of (\ref{eq: 18}) we
have \begin{equation}
\int_{-f_{1}-f_{2}...-f_{\Delta-1}}^{\infty}\rho(f_{\Delta})df_{\Delta}=e^{-1}\; e^{(f_{1}+f_{2}+\ldots+f_{\Delta-1)}/x}.\label{last}\end{equation}
 Since $\rho(f)e^{f/x}=e^{-1}/x$ we end up with 

$\mbox{Prob}(\Delta)=e^{-\Delta}\int_{-1}^{0}df_{1}\int_{-1}^{-f_{1}}df_{2}\int_{-1}^{-f_{1}-f_{2}}df_{3}\ldots$ 

\vskip -.1in

\begin{equation}
..\int_{-1}^{-f_{1}-f_{2}\ldots-f_{\Delta-2}}df_{\Delta-1}.\label{eq:y}\end{equation}

\vskip.05in

\noindent For simplicity we have put $x_{c}=1$, since the quantity
$x_{c}$ simply determines the scale of the steps, and here it is
only relative step lengths that matters.

In Appendix A we have evaluated the expression (\ref{int}), with
the result\begin{equation}
\mbox{Prob}(\Delta)=\frac{e^{-\Delta}\,\Delta^{\Delta-1}}{\Delta!}\simeq\frac{1}{\sqrt{2\pi}}\;\Delta^{-3/2},\label{Pcrit}\end{equation}
 and also shown that these probabilities satisfy \begin{equation}
\sum_{\Delta=1}^{\infty}\mbox{Prob}(\Delta)=1.\label{PRW}\end{equation}

\begin{center}\includegraphics[%
  width=2.4in,
  height=2.1in]{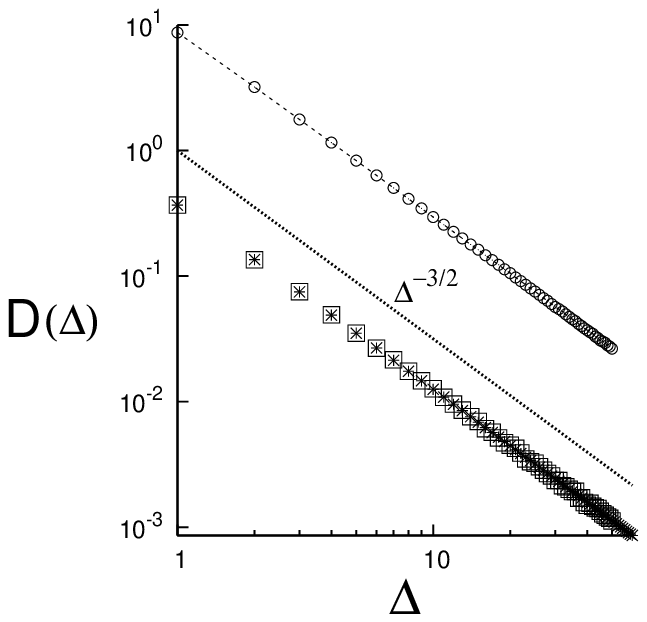}\end{center}

{\footnotesize FIG. 5.} {\footnotesize Distribution of first bursts
(squares) and total bursts (circles) for the critical strength distribution
(\ref{uniform}) with $x_{0}=x_{c}$. The simulation results are based
on $10^{6}$ samples with $N=80000$ fibers. The `star' symbol stands
for the analytic result (\ref{Pcrit}). \par{}}{\footnotesize \par}

\vskip.1in

The result (\ref{PRW}) is strictly applicable only in the limit $N\rightarrow\infty$;
for finite $N$ the sum has to be slightly less than unity. For one
thing $\Delta$ cannot exceed $N$, and breaking off the sum at $\Delta=N$
would decrease the sum by an amount of order $N^{-1/2}$. In reality,
the sum deviates from unity by an amount of order $N^{-1/3}$ (see
the following section).

\vskip.2in

The simulation results in Fig. 5 are in excellent agreement with the
distribution (\ref{Pcrit}). At the completion of a burst the force,
i.e., the excursion of the random walk, is larger than all previous
values. Therefore one may use this point as a new starting point to
find, by the same calculation, the distribution of the next burst,
etc. Consequently the complete burst distribution is essentially $\propto\Delta^{-3/2}$
as expected. In the next section we study the burst distribution at
criticality in more detail, in particular its dependence upon the
bundle size $N$.

\subsection{Finite-size effects at criticality}

When the fiber bundle is sub-critical the average number of bursts
of a given length will be proportional to the bundle size $N$. When
the bundle is critical this is no longer so. Each burst will produce
a non-negligible weakening of the bundle, so that the bundle will
be slightly more supercritical. Then the probability of a total breakdown
will increase, and the probability of a burst of finite length decreases.

To study this quantitatively we therefore have to specify not only
the size $\Delta$ of a burst, but also if it is the first burst after
starting, the second, the third, etc. Let $P_{n}(\Delta)$ be the
number of bursts of size $\Delta$ that occur as the $n$th burst.
If we start precisely at criticality, we have already calculated \begin{equation}
P_{1}(\Delta)=\frac{e^{-\Delta}\Delta^{\Delta-1}}{\Delta!},\end{equation}
 by Eq.(\ref{Pcrit}). We will in particular study how the probability
decreases with $n$, so we form the ratios \begin{equation}
R_{n}(\Delta)=\frac{P_{n}(\Delta)}{P_{1}(\Delta)}.\label{ratios}\end{equation}

We start by investigating the $\Delta$-dependence of these ratios,
and for simplicity we work with the critical uniform threshold distribution
throughout this subsection. In Fig.\ 6 we have plotted $R_{n}(\Delta)$
versus $\Delta$. The ratios (\ref{ratios}) depend upon $n$, but
surprising enough we cannot detect any systematic dependence on $\Delta$.
We may therefore obtain the dependence upon $n$ and $N$ by sticking
to one fixed $\Delta$; for simplicity we take $\Delta=1$.

\noindent \vspace*{1cm}

\begin{center}\includegraphics[%
  width=3in,
  height=3in]{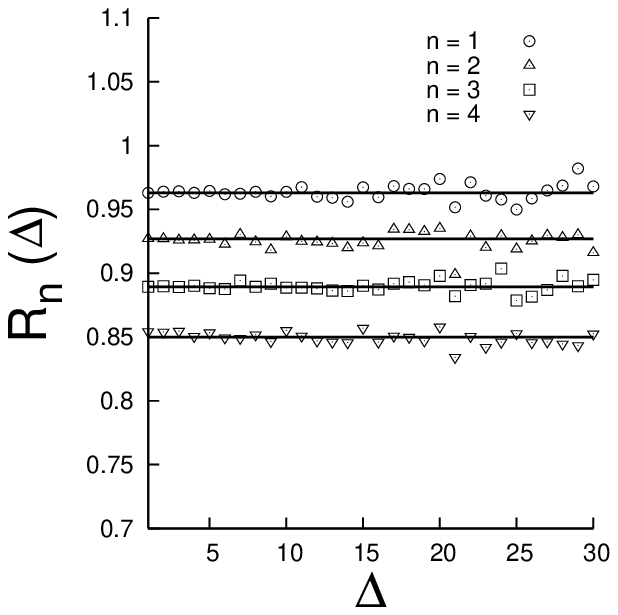}\end{center}

{\footnotesize FIG.\ 6. Simulation results for the $\Delta$-dependence
of the ratios (\ref{ratios}). We have used the uniform strength distribution
at criticality. The figure is based on $10^{7}$ samples with} {\small $N=40000$
fibers}. {\small The straight lines are the average values.}{\small \par}

\vskip.1in

In Fig.\ 7 we have plotted $R_{n}(1)$ for four different values
of $N$ and for $n=2,3,4$ and $5$. The figure shows that $R_{n}(1)$
for each value of $n$ apparently depends linearly on $N^{-1/3}$
and that $1-R_{n}(1)$ apparently is proportional to $n-1$. Empirically
the data can reasonably well be represented by \begin{equation}
R_{n}(1)=1-1.27\;(n-1)N^{-1/3}.\label{R-N}\end{equation}
 More generally we may assume that the linear function is a limiting
form of a more general function: \begin{equation}
R_{n}(1)=F(x),\hspace{1cm}\mbox{with}\hspace{2mm}x=(n-1)N^{-1/3},\label{Fx}\end{equation}
 where $F(0)=1$ and $F(x)\simeq1-1.27x$ for small $x$. For large
$x$ we expect $F(x)$ to approach zero.

If (\ref{Fx}) is correct we should have a data collapse onto the
single curve $F(x)$. Fig.\ 8 shows that the data collapse works
well. In order to test the function $F(x)$ beyond its initial linear
behavior, we have added a few points with larger values of $x$. In
addition to the results of Fig.\ 7 (A) we have obtained results for
$n=10,20$ and $30$, with $N=5000$, $N=10000$ and $N=80000$ (Fig.
7 (B)). 

\begin{flushright}\includegraphics[%
  width=6cm,
  height=6cm]{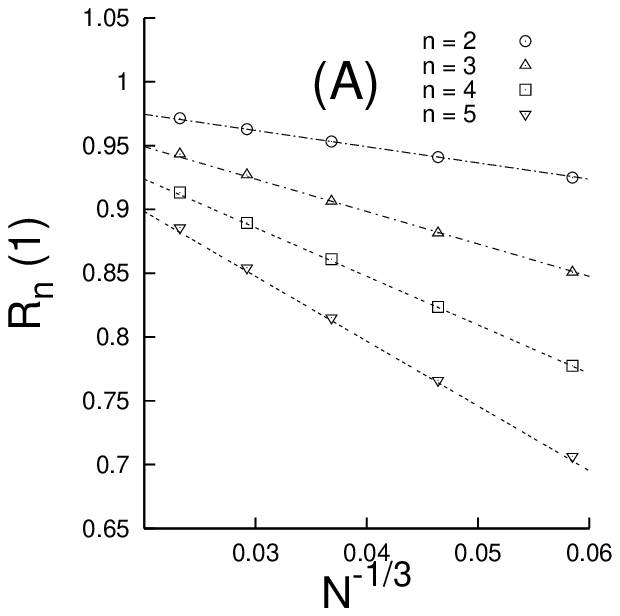}\end{flushright}

\begin{flushleft}\includegraphics[%
  width=6cm,
  height=6cm]{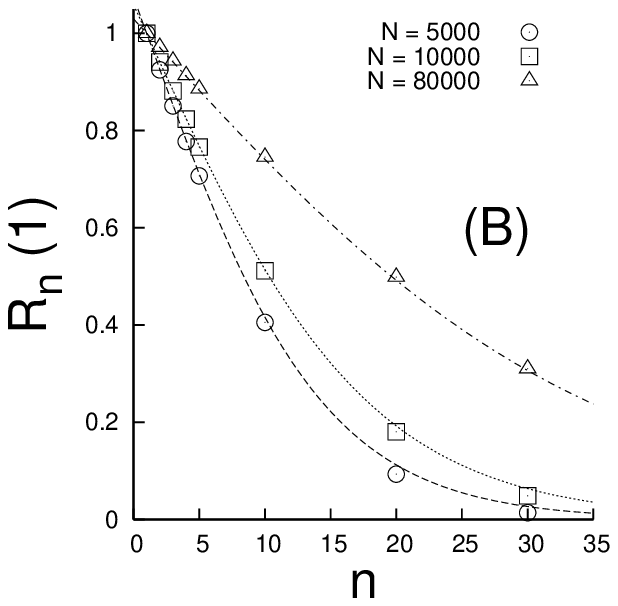}\end{flushleft}

\vskip.1in

{\footnotesize FIG.\ 7. Simulation results for the ratios $R_{n}(1)$
as function of $N^{-1/3}$ (A) and as function of $n$ (B). The dotted
lines represent the functional forms (\ref{R-N}) (in A) and (\ref{emp})
(in B). The results are based on $10^{7}$ samples, except for $N=80000$
($10^{6}$ samples).}{\footnotesize \par}

\vskip.1in

$F(x)$ is seen to decrease towards zero for increasing $x$, and
the empirical expression \begin{equation}
F(x)\simeq\frac{2}{1+e^{2.54x}}\label{emp}\end{equation}
 seems to represent the data in Fig. 8 very well.

The relation (\ref{Fx}) implies some interesting consequences. Let
us first consider the total number of burst of a given size: {\scriptsize \begin{eqnarray}
D(\Delta) & = & \sum_{n}P_{n}(\Delta)=P_{1}(\Delta)\sum_{n}R_{n}(\Delta)\simeq P_{1}(\Delta)\sum_{n}R_{n}(1)\nonumber \\
 & = & P_{1}(\Delta)\sum_{n}F[(n-1)N^{-1/3}]\simeq P_{1}(\Delta)\int_{1}^{\infty}F[(n-1)N^{-1/3}]\; dn\nonumber \\
 & = & P_{1}(\Delta)N^{1/3}\int_{0}^{\infty}F(x)\; dx.\label{toto}\end{eqnarray}
} We have used that $R_{n}(\Delta)$ is essentially independent of
$\Delta$ (Fig. 6), and that due to smallness of $N^{-1/3}$ we may
replace summation by integration.

\begin{center}\includegraphics[%
  width=2.5in,
  height=2.5in]{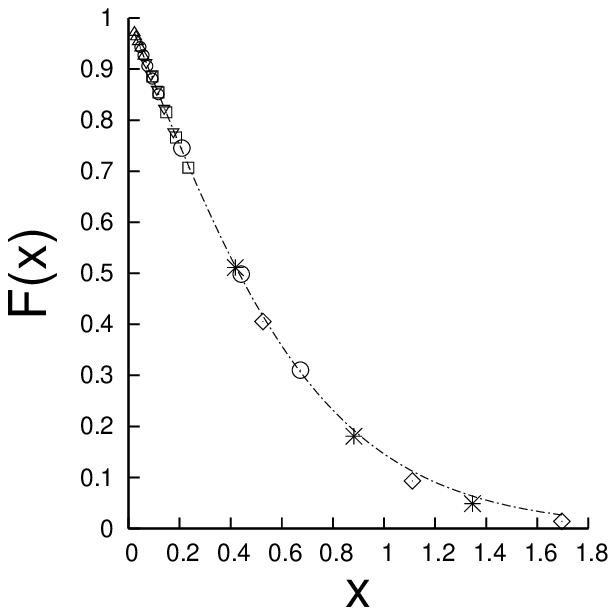}\end{center}

{\footnotesize FIG.\ 8. The data collapse onto a single curve $F(x)$
(dotted line represents Eq. 28) where $x=(n-1)\times N^{-1/3}$.}
{\small We have taken $N=5000,10000,20000,40000,80000$ and $n=2,3,4,5,10,20,30$.
Averages are taken over $10^{7}$ samples,} {\footnotesize except
for $N=80000$ ($10^{6}$ samples)}{\small .}{\small \par}

\vskip.1in

The conclusion is that the total number of bursts should scale as
$N^{1/3}$. The simulation results presented in Table 1 are in excellent
agreement with this $N$-dependence.

\begin{center}Table 1\end{center}

\begin{center}\begin{tabular}{|c|c|c|c|c|c|}
\hline 
$N$&
 $5000$&
 $10000$&
 $20000$&
 $40000$&
 $80000$\tabularnewline
\hline
$N^{-1/3}D(1)/P(1)$&
 $0.545$&
 $0.546$&
 $0.546$&
 $0.550$&
 $0.550$\tabularnewline
\hline
$N^{-1/3}D(2)/P(2)$&
 $0.546$&
 $0.549$&
 $0.550$&
 $0.547$&
 $0.552$\tabularnewline
\hline
$N^{-1/3}D(3)/P(3)$&
 $0.542$&
 $0.550$&
 $0.548$&
 $0.547$&
 $0.546$\tabularnewline
\hline
$N^{-1/3}D(4)/P(4)$&
 $0.538$&
 $0.548$&
 $0.544$&
 $0.548$&
 $0.546$\tabularnewline
\hline
$N^{-1/3}D(5)/P(5)$&
 $0.542$&
 $0.545$&
 $0.546$&
 $0.548$&
 $0.550$\tabularnewline
\hline
$N^{1/3}D$(failure)&
 $1.2154$&
 $1.2319$&
 $1.2346$&
 $1.2293$&
 $1.2307$ \tabularnewline
\hline
\end{tabular}\end{center}

\vskip.1in

According to (\ref{toto}) the numbers in the table should all be
the same (or nearly the same) and represent the integral of $F$.
The integral of the empirical representation (\ref{emp}) of $F(x)$
equals $0.546$, in close agreement with the results in the table.

We have also recorded the number $D(\mbox{failure})$ of \textit{immediate}
failures of the fiber bundle (i.e., with no finite bursts at all).
We have presented the numbers in Table 1. The number of immediate
failures decreases with increasing $N$ as $N^{-1/3}$. The reason
for the decrease is that in a large bundle it is more probable to
find a fiber sufficiently strong to prevent immediate failure.

It remains a challenge to derive these finite-size scaling results
analytically.

\section{Burst avalanches in the fuse model}

\begin{center}\includegraphics[%
  width=2.5in,
  height=2.5in]{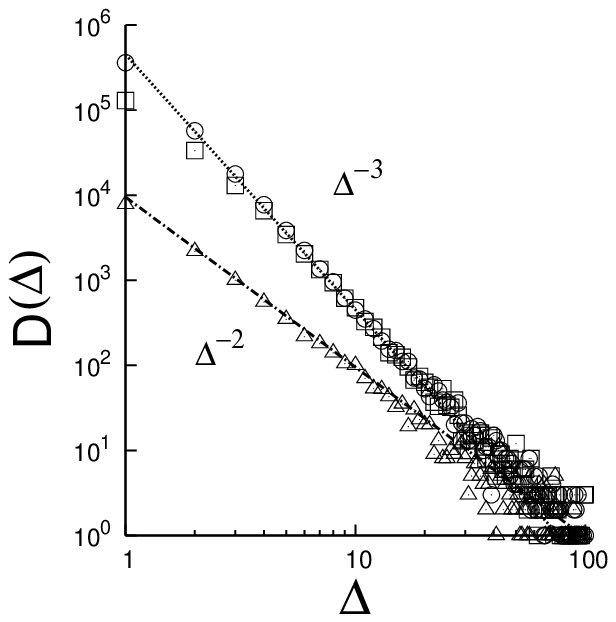}\end{center}

{\footnotesize FIG. 9.} {\footnotesize Burst distribution in the fuse
model: System size is $100\times100$ and averages are taken for $300$
samples . On the average, catastrophic failure sets in after $2097$
fuses have blown. The circles denote the burst distribution measured
throughout the entire breakdown process. The squares denote the burst
distribution based on bursts appearing after the first $1000$ fuses
have blown. The triangles denote the burst distribution after $2090$
fuses have blown. The two straight lines indicate power laws with
exponents $\xi=3$ and $\xi=2$, respectively. }{\footnotesize \par}

\vskip.1in

Let us test the crossover phenomenon in a more complex situation than
for fiber bundles. We have studied burst distributions in the fuse
model \cite{book-1}. It consists of a lattice in which each bond
is a fuse, i.e., an ohmic resistor as long as the electric current
it carries is below a threshold value. If the threshold is passed,
the fuse burns out irreversibly. The threshold $t$ of each bond is
drawn from an uncorrelated distribution $p(t)$. The lattice is placed
between electrical bus bars and an increasing current is passed through
it. Numerically, the Kirchhoff equations are solved with a voltage
difference between the bus bars set to unity. The ratio between current
$i_{j}$ and threshold $t_{j}$ for each bond $j$ is calculated and
the bond having the largest value, $\max_{j}(i_{j}/t_{j})$ is identified
and subsequently irreversibly removed. The lattice is a two-dimensional
square one placed at $45{}^{\circ}$ with regards to the bus bars.
The threshold distribution is uniform on the unit interval. All fuses
have the same resistance. The burst distribution follows the power
law (\ref{powerlaw}) with $\xi=3$, which is consistent with the
value reported in recent studies \cite{fuse-94,fuse-05}. We show
the histogram in Fig. 9. With a system size of $100\times100$, $2097$
fuses blow on the average before catastrophic failure sets in. When
measuring the burst distribution only after the first $2090$ fuses
have blown, a different power law is found, this time with $\xi=2$.
After $1000$ blown fuses, on the other hand, $\xi$ remains the same
as for the histogram recording the entire failure process, see Fig.
9. Zapperi et al [19], who study the fuse model on the diamond and the
 triangular lattices, find significant variation with the lattice type.  
Their exponent values for the diamond lattice are 2.75 and 1.90, not very 
different from the values 3.0 and 2.0 in our Fig. 9.

In Fig. 10, we show the power dissipation $E$ in the network as a
function of the number of blown fuses. The dissipation is given as
the product of the voltage drop across the network $V$ times the
total current that flows through it. In Fig. 11, we show the power
dissipation as a function of the total current. The breakdown process
starts by following the lower curve, and follows the upper curve returning
to the origin. It is interesting to note the linearity of the unstable
branch of this curve. In Fig. 12, we record the avalanche distribution
for power dissipation, $D_{d}(\Delta)$. Recording, as before, the
avalanche distribution throughout the entire process and recording
only close to the point at which the system catastrophically fails,
result in two power laws, with exponents {\small $\xi=2.7$} and {\small $\xi=1.9$,}
respectively. It is interesting to note that in this case there is
not a difference of unity between the two exponents. The power dissipation
in the fuse model corresponds to the stored elastic energy in a network
of elastic elements. Hence, the power dissipation avalanche histogram
would in the mechanical system correspond to the released energy.
Such a mechanical system would serve as a simple model for earthquakes.

\vskip.4in

\begin{center}\includegraphics[%
  width=2.5in,
  height=2.3in]{fig10.eps}\end{center}

{\small FIG. 10. Power dissipation $E$ as a function of the number
of broken bonds in the fuse model. The system size and number of samples
are the same as in Fig. 9.}{\small \par}

\vskip.1in
The Gutenberg-Richter law \cite{book-2,book-3,book-4} relating the
frequency of earthquakes with their magnitude is essentially a measure
of the elastic energy released in the earth's crust, as the magnitude
of an earthquake is the logarithm of the elastic energy released.
Hence, the power dissipation avalanche histogram $D_{d}(\Delta)$
in the fuse model corresponds to the quantity that the Gutenberg-Richter
law addresses in seismology. Furthermore, the power law character
of $D_{d}(\Delta)$ is consistent with the form of the Gutenberg-Richter
law. It is then intriguing that there is a change in exponent {\small $\xi$}
also for this quantity when failure is imminent. 

\vskip.3in

\begin{center}\includegraphics[%
  width=2.4in,
  height=2.1in]{fig11.eps}\end{center}

{\small FIG. 11. Power dissipation $E$ as a function of the total
current $I$ flowing in the fuse model. The system size and number
of samples are the same as in Fig. 9.}{\small \par}

\vskip.4in

\begin{center}\includegraphics[%
  width=2.4in,
  height=2.1in]{fig12.eps}\end{center}

\noindent {\small FIG. 12. The power dissipation avalanche histogram}
$D_{d}(\Delta)$ {\small in the fuse model. The slopes of the two
straight lines are $-2.7$ and $-1.9$, respectively. The circles
show the histogram of avalanches recorded through the entire process,
whereas the squares show the histogram recorded only after $2090$
fuses have blown. The system size and number of samples are the same
as in Fig. 9.}{\small \par}

\section{Concluding remarks}

We have studied the avalanche distribution $D(\Delta)\propto\Delta^{-\xi}$
in the fiber bundle model, and have shown analytically that close
to complete breakdown it exhibits a crossover behavior between two
power laws with exponents $\xi=5/2$ and $\xi=3/2$. This crossover
behavior is universal in the sense that, under mild assumptions, it
does not depend on the statistical distribution of the thresholds.
In the critical situation an argument based on a unbiased unsymmetrical
random-walk scenario explains the exponent $\xi=3/2$. Near criticality
the avalanche distribution depends on the system size in a nontrivial
way. For this case we present quantitative results that may be summarized
by a finite-size scaling function {[}Eq. \ref{emp}{]}. 

The crossover behavior is not limited to the fiber bundle model. We
show numerically that the same crossover phenomenon occurs in the
two-dimensional fuse model. The exponents are different, though, $\xi=2$
near breakdown and $\xi=3$ away from it. For this fuse model the
power dissipation avalanches show a crossover, with power law exponents
{\small $\xi=2.7$} and {\small $\xi=1.9$}. Such crossovers signal
that catastrophic failure is imminent, and has therefore a strong
potential as a useful detection tool. Some of the present results
have already been published as a letter \cite{PHH-05}. 

\vskip.2in

\textbf{Acknowledgment}

S. P. thanks the Research Council of Norway (NFR) for financial support
through Grant No. 166720/V30.

\vskip.2in

\begin{center}\textbf{APPENDIX A: PROOF OF Eq.(\ref{Pcrit})}\end{center}

We evaluate here the multiple integral in Eq.\ (\ref{int}) {\scriptsize \begin{equation}
\mbox{Prob}(\Delta)=e^{-\Delta}\int_{-1}^{0}df_{1}\int_{-1}^{-f_{1}}df_{2}\int_{-1}^{-f_{1}-f_{2}}df_{3}\ldots\int_{-1}^{-f_{1}-f_{2}\ldots-f_{\Delta-2}}df_{\Delta-1}.\label{int2}\end{equation}
} We introduce the new variables \begin{eqnarray}
y_{1} & = & -f_{1}\nonumber \\
y_{2} & = & -f_{1}-f_{2}\nonumber \\
... & = & ...\nonumber \\
y_{\Delta-1} & = & -f_{1}-f_{2}-\ldots-f_{\Delta-1},\end{eqnarray}
 satisfying \begin{eqnarray}
0\leq & y_{1} & \leq1\nonumber \\
0\leq & y_{2} & \leq1+y_{1}\nonumber \\
0\leq & y_{i} & \leq1+y_{i+1}\hspace{15mm}i=2,3,\ldots,\Delta-1.\end{eqnarray}
 Then {\scriptsize \begin{equation}
\mbox{Prob}(\Delta)=e^{-\Delta}\int_{0}^{1}dy_{1}\int_{0}^{1+y_{1}}dy_{2}\int_{0}^{1+y_{2}}\ldots\int_{0}^{1+y_{\Delta-2}}dy_{\Delta-1}.\end{equation}
} Defining $V_{0}(y)=1$, and \begin{equation}
V_{d}(y)=\int_{0}^{1+y}\; V_{d-1}(z)\; dz,\label{iteration}\end{equation}
 we have \begin{equation}
\mbox{Prob}(\Delta)=e^{-\Delta}\, V_{\Delta-1}(0).\label{Prob}\end{equation}
 Equation (\ref{iteration}) can be solved by iteration. By calculating
the first polynomials $V_{n}(y)$ one is led to assume \begin{equation}
V_{d-1}(y)=\frac{1}{d!}\,\sum_{i=1}^{d-1}\, d^{d-i-1}\left(\begin{array}{c}
d-1\\
i\end{array}\right)(i+1)y^{i}.\label{V}\end{equation}
 Suppose this is valid up to some value of $d-1$. Then use (\ref{iteration})
to compute $V_{d}$. The integration is trivial, leaving \begin{eqnarray}
V_{d}(y) & = & \frac{1}{d!}\,\sum_{i=0}^{d-1}\left(\begin{array}{c}
d-1\\
i\end{array}\right)d^{d-i-1}\,(1+y)^{i+1}\nonumber \\
 & = & \frac{1}{d!}\,\sum_{i=0}^{d-1}\sum_{m=0}^{i+1}\left(\begin{array}{c}
d-1\\
i\end{array}\right)\left(\begin{array}{c}
i+1\\
m\end{array}\right)d^{d-i-1}\, y^{m}\nonumber \\
 & = & \frac{1}{d!}\sum_{m=0}^{d}\sum_{i=m-1}^{d-1}\left(\begin{array}{c}
d-1\\
i\end{array}\right)\left(\begin{array}{c}
i+1\\
m\end{array}\right)d^{d-i-1}\, y^{m}\nonumber \\
 & = & \frac{1}{d!}\,\sum_{m=0}^{d}S(m)\; y^{m},\label{Vd}\end{eqnarray}
 with \begin{equation}
S(m)=\sum_{i=m-1}^{d-1}\left(\begin{array}{c}
d-1\\
i\end{array}\right)\left(\begin{array}{c}
i+1\\
m\end{array}\right)d^{d-i-1}.\end{equation}
 Since \begin{equation}
\left(\begin{array}{c}
i+1\\
m\end{array}\right)=\left(\begin{array}{c}
i\\
m-1\end{array}\right)+\left(\begin{array}{c}
i\\
m\end{array}\right),\end{equation}
 and \begin{equation}
\left(\begin{array}{c}
a\\
b\end{array}\right)=0\hspace{10mm}\mbox{for }b<0\mbox{ or }b>a,\end{equation}
 we may write \begin{equation}
S(m)=\sum_{i=0}^{d-1}\left(\begin{array}{c}
d-1\\
i\end{array}\right)\left[\left(\begin{array}{c}
i\\
m\end{array}\right)+\left(\begin{array}{c}
i\\
m-1\end{array}\right)\right]\; d^{d-i-1}.\label{S}\end{equation}
 To evaluate $S(m)$ we differentiate the binomial expression \begin{equation}
\sum_{i=0}^{d-1}\left(\begin{array}{c}
d-1\\
i\end{array}\right)\; x^{1}=(1+x)^{d-1}\end{equation}
 $m$ times with respect to $x$: {\scriptsize \begin{equation}
\sum_{i=0}^{d-1}\left(\begin{array}{c}
d-1\\
i\end{array}\right)i(i-1)\ldots(i-m+1)\; x^{i-m}=(d-1)(d-2)\ldots(d-m)(1+x)^{d-1-m},\end{equation}
} or \begin{equation}
\sum_{i=0}^{d-1}\left(\begin{array}{c}
d-1\\
i\end{array}\right)\left(\begin{array}{c}
i\\
m\end{array}\right)m!\, x^{i-m}=\frac{(d-1)!}{(d-m-1)!}\,(1+x)^{d-1-m}.\end{equation}
 Putting now $x=1/d$, and multiplying both sides of the equation
by $d^{d-1-m}/m!$, we obtain {\footnotesize \begin{equation}
s(m)\equiv\sum_{i=0}^{d-1}\left(\begin{array}{c}
d-1\\
i\end{array}\right)\left(\begin{array}{c}
i\\
m\end{array}\right)d^{d-i-1}=\frac{(d-1)!}{(d-m-1)!m!}\,(1+d)^{d-1-m}.\end{equation}
} By Eq.(\ref{S}) then \begin{equation}
S(m)=s(m)+s(m-1)=\left(\begin{array}{c}
d\\
m\end{array}\right)(1+d)^{d-m}(1+m)/(1+d).\end{equation}
 Finally Eq.(\ref{Vd}) gives \begin{equation}
V_{d}(y)=\frac{1}{(1+d)!}\,\sum_{m=0}^{d}\left(\begin{array}{c}
d\\
m\end{array}\right)(d+1)^{d-m}(1+m)\, y^{m}.\end{equation}
 Since this is in accordance with the assumption (\ref{V}), and since
(\ref{V}) is correct for $d=2$, the induction proof works.

Finally, using (\ref{Prob}), the probability we seek is \begin{equation}
\mbox{Prob}(\Delta)=e^{-\Delta}\; V_{\Delta-1}(0)=\frac{e^{-\Delta}\;\Delta^{\Delta-1}}{\Delta!},\end{equation}
 which is Eq.(\ref{Pcrit}) in the main text.

Let us also sum these probabilities over all burst lengths, {\footnotesize \begin{equation}
S=\sum_{\Delta=1}^{\infty}\frac{e^{-\Delta}\Delta^{\Delta-1}}{\Delta!}=\sum_{\Delta=1}^{\infty}\frac{e^{-\Delta}}{\Delta!}\left[\frac{d^{\Delta-1}}{dS^{\Delta-1}}\; e^{S\Delta}\right]_{S=0}.\end{equation}
} We may now appeal directly to the theorem of Lagrange \cite{WW-1958}
to conclude that the sum satisfies the equation \begin{equation}
S=e^{-1}e^{S}.\label{eq:}\end{equation}
 Since $S\, e^{-S}$ is always less or equal to $e^{-1}$ for nonnegative
$S$, we must have \begin{equation}
S=1,\label{eq:}\end{equation}
 which is Eq.( \ref{PRW}) in the main text.

\end{document}